\documentclass[prd,superscriptaddress,nofootinbib,
preprint]{revtex4}
\usepackage{graphicx}
\usepackage[usenames,dvipsnames]{color}
\usepackage{amsmath,amssymb}
\usepackage{mathtools}
\usepackage[colorlinks]{hyperref}
\usepackage{float}

\newcommand{\be}{\begin{eqnarray}}                                             
\newcommand{\ee}{\end{eqnarray}}
\newcommand{\nn}{\nonumber}

\newcommand{\noplus}{}

\newcommand{\tmop}[1]{\ensuremath{\operatorname{#1}}}
\begin{document}
	\title{  Compact space and infrared behavior of the effective QCD}
		
	\author{Haresh Raval}
	\email{har@physics.iitd.ac.in} 
	\affiliation{Department of Physics, Indian Institute of Technology
		Delhi, New Delhi-110016, India}
	
	\begin{abstract}
  We aim to  investigate the infrared regime in the effective  theory  of the new quadratic gauge in the physical compact space   by defining it on  4-sphere. Abelian dominance  is characterized by off-diagonal gluons acquiring dynamical masses and it hints at  existence of confinement. We do observe Abelian dominance in the ghost condensed vacuum  of this theory  on 4-sphere. Along with this observation, we find  an unusual result that mass of an off-diagonal gluon on 4-sphere is position dependent  in this theory as a consequence of the curved  geometry.  This suggests that  the  curvature  of 4-sphere  does not change the infrared behaviour of present theory on 4-sphere  from that of the theory with the same quadratic gauge in Euclidean spacetime as Abelian dominance is observed  in Euclidean spacetime too while it has caused   mass of an off-diagonal gluon on  4-sphere to be position dependent. The effective action in the confined phase on 4-sphere and that in the 4-dim Euclidean space are found to be identical in a profound outcome.
  	\end{abstract}
  \maketitle
	\section{Introduction}
The   studies of  gauge theory on the compact manifold have been of great theoretical interest   with physical relevance motivated by experimental circumstances, (see e.g., Ref.~\cite{@} and refs therein). Most such studies are on lattice and not analytical whereas we here attempt an analytical study.  The  4-sphere is a compact space therefore, an effective theory on 4-sphere provides a model to study consequences of compactness of  physical space on phenomena in QCD. Thus,  the present theory on 4-sphere is purposed to be a model to investigate confinement in the finite physical volume analytically.
	The idea of formulating gauge theories on 4-sphere and   hypersphere in general commences with a study of massless Euclidean QED (quantum electrodynamics) on a hypersphere in 5-dimensional Euclidean space\cite{1}. It was argued there that  compactification
	of space-time leads to infrared-finiteness of this theory. Thereafter manifestly $O(n)$-covariant formulation of gauge theories on a hypersphere
	have been considered at a few occasions in different contexts\cite{2,3,4,5,5a,5a',5b}. For example, the manifestly
	$O(n)$-covariant formulation was reconstructed using conformal Killing vectors in Ref.\cite{5a'}.  To the best of our knowledge, non perturbative inspection in QCD on 4-sphere has never been carried out. 
	We are required to introduce a suitable gauge-fixing in
	this theory to analyze the underlying field theoretical properties of a gauge theory.
	 In the study of massless Euclidean QED on a hypersphere,
     the  following gauge condition was deliberately chosen such that analysis retains a 
     manifest  $O(n)$-covariance\cite{1}, 
     \be \label{1}
      (r_\alpha \partial_\beta-r_\beta \partial_\alpha) \bar{A}_\beta =  \bar{A}_\alpha; 
       \text{  $r_\alpha$  is cartesian coordinate, $r_\alpha r_\alpha =1$, $\alpha = 1,2,...,n$. } \ee
       Here $ \bar{A}$ is the gauge field on the hypersphere.  
     With this condition,  extension of the analysis 
	  to the non abelian gauge theory is straight forward. However,
	the gauge   is  inconsistent  for  gauge fixing procedure on sphere as it has got an extra free index. The gauge given in Eq.\eqref{2} rectifies the flaw since it does not have any free index and is equivalent of the gauge in Eq.\eqref{1}\cite{5a}, 
	\be\label{2}
 r_\alpha (r_\alpha \partial_\beta-r_\beta \partial_\alpha) \bar{A}_\beta = r_\alpha iM_{\alpha \beta} \bar{A}_\beta = 0.
	\ee
	This gauge despite being equivalent to  that in Eq.\eqref{1} 
 has more similar form to $\partial_\mu A_\mu =0$ than the former.  
	  One can fairly expect that compactification
	  regularizes infrared divergences in QCD also. For this reason,   QCD
	 on 4-sphere $\mathbb{S}^{4}$ would be more appropriate for analyzing properties of QCD at a low-energy region.  This provides us the motive to apply the proposed gauge on $\mathbb{S}^{4}$ as it has already shown nice low-energy attributes in the flat 4-dim space.
	 
	 The quadratic gauge
	  has been extensively studied in various settings recently\cite{6,7,8,9,10,11,epl} and  found to have substantial implications in non-perturbative sector of QCD. It is given  as follows
	 	\begin{align} \label{eq:0}
	 H^a [ A^{\mu} ( x) ] =
	 A^a_{\mu} ( x) A^{\mu a} ( x) = f^a ( x) ; \  \text{  for each $a$, }
	 \end{align}
	 where $f^a(x)$ is an arbitrary function of $x$. The Faddeev-Popov action in the flat space is given as
	 	\begin{eqnarray} \label{eq:Leff}
	 \mathcal{L}_{Q}	 = - \frac{1}{4} F^a_{\mu \nu} F^{\mu \nu a}
	 \noplus + \frac{\zeta}{2 }  F^{a2}+ F^a A^a_{\mu} A^{\mu a} -2 \overline{c^a}
	 A^{\mu a} ( D_{\mu} c)^a  ,
	 \end{eqnarray}
	where  the field strength $F^a_{\mu \nu}= \partial_{\mu}A^a_{\nu}(x)- \partial_{\nu}A^a_{\mu}(x)-g 
	f^{abc} A^b_{\mu}(x)A^c_{\nu}(x)$, the $F^a$ are a set of auxiliary fields,  $c, \bar{c}  $ are ghost and anti-ghost fields respectively,  $\zeta$ is an arbitrary gauge fixing parameter and $(D_{\mu} c)^a = \partial_\mu c^a - g f^{a b c} A_\mu^b c^c$.   The summation over indices $a$, $b$ and $c$ each runs independently over $1$ to $N^2-1$ in Eq.\eqref{eq:Leff}.
	 
	Now we present the plan of the paper.  Next section   reviews  Yang-Mills theory on a general hypersphere and   the gauge fixing procedure  for condition in Eq.\eqref{2} using BRST invariance as the first principle. In section III however we limit our selves to 4-sphere in order to render the study relevant to physical situation  i.e., the compact space study of the phenomenon under consideration.  We develop the effective theory in the quadratic gauge on  4-sphere with the help of  Faddeev-Popov procedure rather than BRST invariance as the first principle and  probe non trivial ghost vacua  in this theory.  In the last section, we conclude this letter.

\section{SU(N) QCD on hypersphere}
Here we review the manifestly $O(n)$ covariant formalism of QCD on $ (n-1)$-dimensional hypersphere, $\mathbb{S}^{n-1}$. The sphere is embedded in the $n$-dimensional Euclidean space $\mathbb{R}^{n}$.    We consider a unit radius $\mathbb{S}^{n-1}$ to bring neatness in mathematical expressions as physics does not depend upon the size.  So we have the constraint, $r_\alpha r_\alpha =1$ with $r_\alpha$ being the cartesian coordinate of the point on $\mathbb{S}^{n-1}$ and $\alpha = 1,2,...,n$. Therefore, the coordinate $r_n$ is dependent variable since $r_n= \pm \sqrt{1- r_\mu r_\mu}$ where  $\mu = 1,2,...,n-1$. For QCD on hypersphere, the propagation of fields remains  transverse to radial direction. Spherical symmetry in the system  preserves the angular momentum, therefore the fundamental operator that governs the dynamics is the  angular momentum operator given as below
\be\label{de92}
       M_{\alpha \beta}   = -i ( r_\alpha \partial_\beta-r_\beta \partial_\alpha);\ \    \partial_\beta \equiv \frac{\partial}{\partial r_\beta},\ \ \alpha, \beta = 1,2,...,n.
       \ee
In particular,
\be\label{de91}
M_{\mu n} = -M_{n \mu}=  i r_n \partial_\mu,
\ee
since $ \frac{\partial}{\partial r_n}\equiv 0$ by definition regardless of  what it operates on as $r_n$ is dependent variable, we shall come to this point again and relax this situation a bit   without disturbing the physics.
The operator $ M_{\alpha \beta}$ satisfies the following Lie algebra
\be\label{al}
[M_{\alpha \beta},M_{\gamma\eta} ] = i(\delta_{\alpha\gamma}M_{\beta\eta} -\delta_{\beta\gamma}M_{\alpha\eta} - \delta_{\alpha\eta}M_{\beta\gamma} +\delta_{\beta\eta}M_{\alpha\gamma}).
\ee
Now we introduce  stereographic coordinates $\{x_\mu, \mu = 1,2,...,n-1\} $ on the hyperplane  $\mathbb{R}^{n-1}$. They map the coordinates for the point on a sphere, $r_\mu$ as 
\be\label{st1} 
r_\mu = \frac{2 x_\mu}{1+x^2},\ \ \  r_n = \frac{1-x^2}{1+x^2},\ \ \  x^2 \equiv x_\mu x_\mu.
\ee
The gauge field on $\mathbb{S}^{n-1}$ is Lie algebra valued i.e., $\bar{A}_\mu= \bar{A}_\mu^a T^a$, where\ $ T^a$ are generators of SU(N) group. The gauge field $\bar{A}$ maps to the gauge field $A$ on the stereo graphic hyperplane as follows\cite{2}
\be\label{st2}
\bar{A}_\mu (r) = \frac{1+x^2}{2} A_\mu(x) -x_\mu x_\nu A_\nu(x), \ \ \bar{A}_n  =-x_\mu A_\mu;\ \ \mu, \nu = 1,2,...,n-1.
\ee
As we are in Euclidean frame we do not bother about upper and lower indices.      The following transversality condition is inherent   in the QCD on hypersphere as it results from innate stereographic projections in Eqs.\eqref{st1},\eqref{st2} 
\be
r_\alpha \bar{A}_\alpha =0 \Rightarrow r_\alpha \bar{A}_\alpha^a=0,\ \  \alpha = 1,2,...,n.
\ee It implies that  the gluon $\bar{A}^a$  lives on the tangent space of the $\mathbb{S}^{n-1}$. Therefore, the equivalence of gauges in Eqs.\eqref{1},\eqref{2} is apparent.

The structure of a gauge transformation encodes the effect of the spatial geometry on the
effective theory. Therefore, the effective theory in the same gauge on different geometries
has different forms.
Let us now turn to the gauge transformation of the gluon on $\mathbb{S}^{n-1}$  as follows
\be \label{3}
\delta \bar{A}^a_\beta =  r_\alpha(r_\alpha D_\beta-r_\beta D_\alpha) \epsilon^a =  r_\alpha i \mathcal{M}_{\alpha \beta }\ \epsilon^a,  \ee
where $\epsilon$ is a parameter of transformation, $D_\beta \epsilon^a =  \partial_\beta \epsilon^a -g f^{abc} \bar{A}_\beta^b \epsilon^c$ is a usual covariant derivative of a parameter. We note that  the  operator in the bracket, $ \mathcal{M}_{\alpha \beta }$ is a covariantized version of the angular momentum operator due to local symmetry. Contracting  Eq.\eqref{3} with $r_\beta$ we get zero which confirms that the infinitesimal gauge transformation is always  tangential to  $\mathbb{S}^{n-1}$.  Thus, the transformed and original field both live on the same tangent space. The Eq.\eqref{3} can be simplified a bit as follows
\be \label{4}
\delta \bar{A}^a_\beta =   ( D_\beta- r_\alpha r_\beta D_\alpha) \epsilon^a, \ \ \text{ since } \ r_\alpha r_\alpha =1.
\ee
The conventional field strength $\bar{F}^a_{\alpha \beta}$ of $\bar{A}$ as defined in the introduction is not  $O(n)$ covariant under the transformation in Eq.\eqref{4} as we see from the expression below
\be\label{5}
\delta \bar{F}_{\alpha\beta} ^a = f^{abc } \bar{F}_{\alpha\beta}^b \epsilon^c+ r_\alpha\left(  D_\beta + \frac{1}{r_n}\delta_{\beta n}\right)r_\mu \partial_\mu \epsilon^a -  r_\beta\left(  D_\alpha + \frac{1}{r_n}\delta_{\alpha n}\right)r_\mu \partial_\mu \epsilon^a.
\ee
We notice that the last two terms are not manifestly covariant. To get rid of such terms, a rank-3 tensor is required. We can deduce the following from Eq.\eqref{5}
 \be
r_\gamma \delta \bar{F}_{\alpha\beta} ^a + r_ \beta \delta \bar{F}_{ \gamma \alpha } ^a   +  r_ \alpha \delta \bar{F}_{ \beta \gamma} ^a =   f^{abc }  (r_\gamma \bar{F}_{\alpha\beta}^b  +    r_\beta \bar{F}_{\gamma \alpha  }^b  +    r_ \alpha   \bar{F}_{ \beta \gamma}^b )\epsilon^c,
\ee
which is covariant manifestly. Therefore we define a rank-3 tensor as 
\be
\bar{F}_{\alpha \beta \gamma }^a = r_\gamma   \bar{F}_{\alpha\beta} ^a + r_ \beta   \bar{F}_{ \gamma \alpha } ^a   +  r_ \alpha   \bar{F}_{ \beta \gamma} ^a.
\ee 
    With the field strength $\bar{F}_{\alpha \beta \gamma }^a $,   the gauge invariant QCD action on $\mathbb{S}^{n-1}$  is given by
    \be \label{8}
    S_{YM} = -\frac{1}{12} \int d \Omega  \ \bar{F}_{\alpha \beta \gamma }^a \bar{F}_{\alpha \beta \gamma }^a.
    \ee
    Here  $d \Omega $ is an invariant measure on $\mathbb{S}^{n-1}$
      which is given in terms
    of the coordinates ($r_\mu$) as follows
   \be 
   d \Omega= \frac{1}{|r_n|} \displaystyle\prod\limits_{\mu=1}^{n-1} d r_\mu.
   \ee
   We can now analyze the effective theory of the condition in Eq.(\ref{2})  using the BRST invariance\cite{new1,new2,new3} as the first principle\cite{5a}.   In this gauge fixing procedure, we write the gauge fixing and ghost term to be added in Eq.\eqref{8} as a total BRST differential as shown below
   \be\label{6}
   S_{GF} +S_{ghost}  = \int d \Omega \ \delta\left[\hat{\bar{c}}^a(r_ \alpha i M_{\alpha\beta}\bar{A}^a_\beta + \frac{\zeta}{2}\bar{B}^a )\right].
\ee  
The overbar indicates fields on sphere.
The BRST transformations are given as below
\be\label{9}
\delta \bar{A}^a_\beta &=&   ( D_\beta- r_\alpha r_\beta D_\alpha) \bar{c}^a =  r_\alpha i\mathcal{M}_{\alpha \beta } \bar{c}^a,\nn\\
\delta \bar{c}^a &=&\frac{1}{2} f^{abc} \bar{c}^b\bar{c}^c,\nn\\
\delta\hat{\bar{c}}^a &=&  \bar{B}^a,\\
\delta \bar{B}^a&=&0.\nn
\ee
We note that only the gauge field has space index and thus only its BRST transformation changes on the hypersphere from that in the flat space. The BRST transformations of the rest are the same as those in the flat space. Nilpotency of the transformation assures the BRST invariance of the resulting effective theory. Using these transformations, we  expand the Eq.\eqref{6} to get
\be\label{7}
 S_{GF} +S_{ghost} =  \int d \Omega \left[\frac{\zeta}{2}(\bar{B}^a )^2 + \bar{B}^ar_ \alpha iM_{\alpha\beta}\bar{A}^a_\beta - \hat{\bar{c}}^a r_ \alpha iM_{\alpha\beta} r_ \gamma i\mathcal{M}_{\gamma\beta} \bar{c}^a\right].
 \ee
We have now all that we need to move on to the  main purpose of the paper.

	\section{Effective theory on 4-sphere and its non perturbative sector}
	In refs.~\cite{6,9}, we have shown that Abelian dominance is observed in the effective quadratic gauge fixed theory in the flat space. 
	Here we limit the discussion to 4 dimensions. We propose the  effective theory in the same   quadratic gauge on 4-sphere and address the issue of confinement in the compact space
	by  analyzing its non perturbative regime.
	 We derive the theory with Faddeev Popov (FP) method rather than BRST invariance as the first principle. The generalization to higher dimensions is obvious. The FP method relies on the FP operator which depends upon the choice of gauge fixing and gauge transformation. As mentioned earlier, the transformation is determined by underlying space hence,  we must choose the transformation on 4-sphere  (Eq.\eqref{3}) in deriving the required FP operator   to  obtain the effective action   in the quadratic gauge  built on the $\mathbb{S}^{4}$ which we get as in Eq.\eqref{28}.  We shall then investigate Abelian dominance by probing mass matrix for gluons on $\mathbb{S}^{4}$ in the ghost condensed phase of this effective theory in which all the ghost condensates are identical. We also find that the necessity to realize this condensate identity within a physical mechanism consistent  on the $\mathbb{S}^{4}$ leads to mass of an off-diagonal gluon on $\mathbb{S}^{4}$ to be position dependent. The  quadratic gauge and the corresponding FP operator $ \Delta_{FP}$ on $\mathbb{S}^{4}$ are respectively as follows     
	\be 
	&&\bar{A}^a_\beta (r)  \bar{A}^a_\beta (r) = f^a(r), \ \ \beta = 1,2,...,5; \ \text{for each $a$ and,}\\
 \Delta_{FP}&=&  \det\left[2\bar{A}^a_\beta\Big( \partial_\beta \delta^{ab}-g f^{acb}\bar{A}_\beta^ c- r_\alpha r_\beta(\partial_\alpha \delta^{ab}-g f^{acb}\bar{A}_\alpha^ c)\Big)\right] .
	\ee  
	Therefore, we get the gauge fixing and the ghost terms as below
	\be\label{10}
	S_{GFQ} +S_{ghostQ} =  \int d \Omega \left[ \frac{\xi}{2}(\bar{B}^a)^2 + \bar{B}^a \bar{A}^a_\beta  \bar{A}_\beta^a-2 \hat{\bar{c}}^a \bar{A}_\beta^a D_\beta \bar{c}^a+2 \hat{\bar{c}}^a \bar{A}_\beta^a  r_\alpha r_\beta D_\alpha \bar{c}^a  \right],
	\ee
		$d\Omega$ is angular measure for 4-sphere.   The inherent transversality gets rid of the last
		ghost term in Eq.\eqref{10}, therefore it actually contains only the first three terms,
		\be\label{28}
		S_{GFQ} +S_{ghostQ} =  \int d \Omega \left[ \frac{\xi}{2}(\bar{B}^a)^2 + \bar{B}^a \bar{A}^a_\beta  \bar{A}_\beta^a-2 \hat{\bar{c}}^a \bar{A}_\beta^a D_\beta \bar{c}^a\right],\ \ \ \text{ since  } \  r_\beta \bar{A}_\beta^a=0.
		\ee
		 The resulting action \be S_{eff}=S_{YM}+S_{GFQ} +S_{ghostQ}, \ee is invariant under the same BRST transformations of Eq.\eqref{9}. Therefore,  the theory is FFBRST\cite{f1, f2, f3} compatible as it is BRST invariant which means that FFBRST technique can be applied to the current theory to connect it to a different effective theory e.g., Lorenz gauge fixed theory on 4-sphere. It is an interesting problem as FFBRST  in the curved geometry is scarcely explored.  The details are however subject of future study.
		
		Before moving further, we like to find the equivalent of the quadratic gauge on the 4-sphere and see whether such equivalent gauge is of some interest.   The 
		transversality condition  
		\be
	 r_\alpha \bar{A}_\alpha^a=0,\ \  \alpha = 1,2,...,5,
		\ee
		gives us the following identity upon differentiation\cite{5a}
		\be\label{11}
		r_\alpha \partial_\beta\bar{A}^a_\alpha = -\bar{A}^a_\beta +\frac{r_\beta}{r_5}\bar{A}_5^a.
		\ee
		Now,
		\be \bar{A}^a_\beta \bar{A}^a_\beta &=& \bar{A}^a_\beta \left[\bar{A}^a_\beta - r_\beta \frac{\bar{A}^a_5}{r_5}\right] \nn\\
		&=& - \bar{A}^a_\beta \ r_\alpha \partial_\beta \bar{A}^a_\alpha \ \ \ \ \ \  \text{since Eq.\eqref{11} }\nn\\ 
	&=&  \bar{A}^a_\beta\  (r_\beta \partial_\alpha - r_\alpha\partial_\beta)  \bar{A}^a_\alpha\ \ \ \ \  \ \   \text{since   $r_\alpha \bar{A}_\alpha^a=0$}\nn\\
	& =&  \bar{A}^a_\beta i M_{\beta \alpha}  \bar{A}^a_\alpha. \label{12}\ \ \ \  \ \  
	\ee 
	The Eq.\eqref{12} in general implies that $i M_{\beta \alpha}  \bar{A}^a_\alpha= \bar{A}^a_\beta + r_\beta W^a $, where $W^a$ is a function of fields $\bar{A}^a_\alpha, \bar{B}^a, \bar{c}^a, \hat{\bar{c}}^a$. We note that the gauge  in Eq.\eqref{1} emerges  as a special case  of equivalence with $W^a=0$.  The Eq.\eqref{12}  also implies that vectors $i M_{\beta \alpha}  \bar{A}^a_\alpha-\bar{A}^a_\beta  $ and $ \bar{A}^a_\beta $ are mutually orthogonal. Therefore,  the vector $i M_{\beta \alpha}  \bar{A}^a_\alpha-\bar{A}^a_\beta  $ is directed radially.   
	
	\subsection{Infrared regime of the theory}
	
	We focus on the ghost Lagrangian which is expanded as follows
	\be\label{gh}
	2 \hat{\bar{c}}^a \bar{A}^{ a}_\beta  D_\beta \bar{c}^a = 2 \hat{\bar{c}}^a \bar{A}^{ a}_\beta \partial_\beta \bar{c}^a -2	gf^{abc}   \hat{\bar{c}}^a {\bar{c}}^c \bar{A}_\beta^b \bar{A}_\beta^a. 
	\ee
	In the desired ghost condensed phase, the first term on the right vanishes as we show in the last paragraph of the present section.
The second  term  is suggestive of the mass matrix for gluons on $\mathbb{S}^{4}$  as follows
 	\begin{equation}
 (M^{ 2})^{a b}_{\tmop{dyn}} = 2 g \displaystyle\sum\limits_{c=1}^{N^2-1}f^{a b c} 
 \langle\hat{\bar{c}}^a \bar{c}^c\rangle.
 \end{equation}
 In the  symmetric state, where all ghost-anti{\small -}ghost 
 condensates are  identical i.e., 
 \begin{equation}\label{13}
 \langle\hat{\bar{c}}^1 \bar{c}^1\rangle = ... =  \langle\hat{\bar{c}}^1 \bar{c}^{N^2-1}\rangle = ... =  \langle\hat{\bar{c}}^{N^2-1} \bar{c}^1\rangle = ... =     
 \langle\hat{\bar{c}}^{N^2-1} \bar{c}^{N^2-1}\rangle = K',
 \end{equation}
 an interesting situation arises. 	
 The mass matrix in this state becomes 
 \begin{equation}\label{mm}
 (M^{ 2})^{a b}_{\tmop{dyn}} = 2 g \displaystyle\sum\limits_{c=1}^{N^2-1}f^{a b c} K',
 \end{equation}
 	which is an anti-symmetric matrix 
 due to the anti-symmetry of the structure constant $f^{a b c}$.
 The resulting mass matrix  has $N(N - 1)$ non-zero eigenvalues only   and thus has 
 nullity $N -1$. The non-zero eigenvalues occur in conjugate pairs. This implies $N(N - 1)$ off-diagonal gluons on sphere have become massive with  $M_{gluon} = \frac{1}{\sqrt{2}}(1\pm i)m$ and $N-1$ diagonal gluons remain massless. Thus in this description, 
 the positive real part i.e., $ Re(M_{gluon}) > 0$,  of the off-diagonal gluon mass makes interactions
 short-ranged. Only the diagonal gluons mediate interactions in the IR limit which are long range, strongly indicating Abelian dominance on the sphere.  Abelian dominance is observed in the effective quadratic gauge fixed theory in the Euclidean space too\cite{6,9}.   Thus, non perturbative behaviour of the quadratic gauge fixed theory on $\mathbb{S}^4$   and in $\mathbb{R}^4$ is  similar.  The form similarity of ghost Lagrangian on $\mathbb{S}^4$  and in $\mathbb{R}^4$ (See Eqs.\eqref{28} and \eqref{eq:Leff}) is particular to this gauge only. The  ghost Lagrangian of Lorenz gauge, for example, on $\mathbb{S}^4$  and in $\mathbb{R}^4$ are not form similar.
 
 		It is imperative to realize the identity in Eq.\eqref{13} within the consistent mechanism on 4-sphere for rendering  the non perturbative consequence of the theory physically relevant. The realization of the identity   within the physical mechanism is not possible through the Coleman Weinberg approach on the curved spacetime  unlike in the flat spacetime. There is  only one way to demonstrate  the identity in the Eq.\eqref{13}   in a physically consistent manner on the curved spacetime  and that is to use two known  results.
    The first is the mapping between ghosts on sphere and corresponding ghosts in the  flat space~\cite{5a}
 \be\label{mmap} 
 \hat{\bar{c}}^a (r) = \frac{1 + x^2}{2} \tilde{c}^a(x), \ \ \bar{c}^a (r) = \frac{1 + x^2}{2} {c^a}(x),
 \ee
 where $\tilde{c}(x), c(x)$ are anti-ghost and ghost fields respectively in the Euclidean space. Second is the 
 following identity for ghosts in the Euclidean spacetime $\mathbb{R}^4$  whose realization  within the Coleman Weinberg mechanism is  given  in \cite{6,9},
 \be\label{id}
  \langle{\tilde{c}}^1 {c}^1\rangle = ... =  \langle{\tilde{c}}^1 {c}^{N^2-1}\rangle = ... =  \langle{\tilde{c}}^{N^2-1} {c}^1\rangle = ... =     
 \langle{\tilde{c}}^{N^2-1} {c}^{N^2-1}\rangle = K,
 \ee
 where $K$ is a constant \cite{6}.
 Putting Eq.\eqref{mmap}  in Eq.\eqref{id}, we  get the  identity in Eq.\eqref{13},
 \be\label{32}
  \langle\hat{\bar{c}}^1 \bar{c}^1\rangle = ... =  \langle\hat{\bar{c}}^1 \bar{c}^{N^2-1}\rangle = ... =  \langle\hat{\bar{c}}^{N^2-1} \bar{c}^1\rangle = ... =     
 \langle\hat{\bar{c}}^{N^2-1} \bar{c}^{N^2-1}\rangle= (\frac{1 + x^2}{2})^2 K = K'.
 \ee
 	Therefore, mass matrix in Eq.\eqref{mm} becomes 
 \be\label{mm'}		(M^{ 2})^{a b}_{\tmop{dyn}} = 2 g \displaystyle\sum\limits_{c=1}^{N^2-1}f^{a b c} K'=  2  g  \big[ \frac{1 + x^2}{2}\big]^2  \Big(\displaystyle\sum\limits_{c=1}^{N^2-1} f^{a b c} K \Big).
 \ee
 Since $K$ is constant of Eq.\eqref{id},  eigenvalues of the matrix in the bracket in  Eq.\eqref{mm'} are constants. We denote them by $m_a^{*2}, a$ is a group index. Here, it is clear that $m_a^{*2}=0$ when $a$ is a diagonal index.  Hence by diagonalizing Eq.\eqref{mm'}, we get  the following expression for the square of mass of a gluon on $\mathbb{S}^4$, $M_{a}^2 $
 \be \label{mr}
 M_{a}^2 =2 g  \big[ \frac{1 + x^2}{2}\big]^2  m_{a}^{*2} = (1 +r_5)^{-2} m_a^2, \ \  m_a^2 \equiv 2g m_{a}^{*2}. 
 \ee
 	Non zero $m_{a}^2$ and thus $M_{a}^2$ are purely imaginary numbers.  $ M_{a}^2=m_{a}^2=0$ for  diagonal gluons. It is further important to note a point  about $m_a^2$. The fact that $K$   in the bracket in Eq.\eqref{mm'} corresponds to  ghosts in Euclidean space as given in Eq.\eqref{id} and that  the  quadratic term in gluons on $\mathbb{S}^4$ relates to quadratic term in gluons in  $\mathbb{R}^4$ only as shown below in Eq.\eqref{al1}  
 clearly establishes that $m_a^2$ in Eq.\eqref{mr} are nothing but squares of  masses of  gluons $A(x)$ in the flat space $\mathbb{R}^4$,
\be\label{al1}
  \bar{A}^a_\beta  \bar{A}_\beta^b (r)&=& (\frac{1 + x^2}{2})^{2}  {A}^a_\mu  {A}_\mu^b(x).
\ee
It is easy to prove the above identity from the mapping in Eq.\eqref{st2}.
 The Eq.\eqref{mr} suggests that mass of an off-diagonal gluon
   on the  4-sphere is position dependent which reflects the consequence of the curved spatial
   geometry on mass in this theory. Specifically, mass of an off-diagonal gluon remains same on a horizontal
   cross section of the  sphere $\mathbb{S}^4$, which is a `circle' $\mathbb{S}^3$  whereas it varies on
   such parallel cross sections. Thus, the curvature influences  mass of an off-diagonal gluon on $\mathbb{S}^4$  without altering   the infrared behaviour of this theory on $\mathbb{S}^4$  from that of the quadratic gauge fixed theory in the $\mathbb{R}^4$. 
   
    Now we show that the  term  $\hat{\bar{c}}^a \bar{A}^{\beta a} \partial_\beta \bar{c}^a$ in Eq.\eqref{gh} disappears in the ghost condensed phase.
 We note that $  \langle{\tilde{c}}^a\frac{\partial}{\partial x^\mu} {c}^a\rangle =0 $~\cite{6}   in the flat space since $\langle{\tilde{c}}^a {c}^c\rangle = const.$ for all $x$. Therefore, using the mappings, we get
 \be\label{@@@}
  \langle{\tilde{c}}^a\frac{\partial}{\partial x^\mu} {c}^a\rangle= \frac{8}{(1 + x^2)^3} \langle \hat{\bar{c}}^a K_{\mu \alpha} \frac{\partial}{\partial r^\alpha}\big[  \frac{2}{(1 + x^2)} \bar{c}^a \big]\rangle =0,
  \ee where $K_{\mu \alpha}$ is a killing vector,
  \be
 K_{\mu \nu}=  \frac{1+x^2}{2} \delta_{\mu \nu} -x_\mu x_\nu, \   K_{\mu n}= -x_\mu.\nn\ee
      In the usual scenario $ \frac{\partial}{\partial r^5}  \equiv 0$ regardless of what it operates on,   which we now relax a bit and  apply as an only proposition the following 
  \be 
   \frac{d}{d r_5} r_5& = &-1. 
  \ee
  The  algebra in Eq.\eqref{al} is almost retained under  this proposition,  i.e.,
    \begin{eqnarray}\label{40}
      &&[M_{\mu \nu},M_{\gamma \eta} ] = i(\delta_{\mu \gamma}M_{\nu \eta} -\delta_{\nu \gamma}M_{\mu \eta} - \delta_{\mu \eta}M_{\nu \gamma} +\delta_{\nu \eta}M_{\mu \gamma});\\ &&\gamma, \eta= 1,...,5, \mu, \nu= 1,...,4,\nn
      \end{eqnarray}
    \be\label{42}
    [M_{\mu 5}, M_{\lambda 5}] = 0.
    \ee
    The only change introduced by the conjecture is that the commutator vanishes in the Eq.\eqref{42}
    which simplifies an algebra marginally, the rest in the Eq.\eqref{40} remain unchanged. Thus, the conjecture  applied does not disturb the underlying physics of the theory. Using this proposition in Eq.\eqref{@@@}, we can verify that 
    \be  \langle \hat{\bar{c}}^a K_{\mu \alpha} \frac{\partial}{\partial r^\alpha}\big[  \frac{2}{(1 + x^2)} \bar{c}^a \big]\rangle  & =&\ 
    \langle \hat{\bar{c}}^a K_{\mu \alpha} \frac{\partial}{\partial r^\alpha}\big[  (1+r_5) \bar{c}^a \big]\rangle =0\nn\\
&    \Rightarrow &K_{\mu \alpha} \langle\hat{\bar{c}}^a \frac{\partial}{\partial r^\alpha}\bar{c}^a\rangle =0 \ \text{ as}\  K_{\mu \alpha}r_\alpha =0\nn\\
&    \Rightarrow & \langle\hat{\bar{c}}^a \frac{\partial}{\partial r^\alpha}\bar{c}^a\rangle =0.\ee
     	Therefore, the  term    $\langle\hat{\bar{c}}^a \bar{A}^{\beta a} \partial_\beta \bar{c}^a\rangle=0$.   Hence, the  effective action in the ghost condensed phase now becomes
    \be\label{''}
    S_{eff}=S_{YM} +  \int d \Omega \Big[\frac{\xi}{2} (\bar{B}^a)^2 +\bar{B}^a \bar{A}^a_\beta  \bar{A}_\beta^a + M_{a}^2 \bar{A}^a_\beta  \bar{A}_\beta^a\Big].
    \ee
 

	\section{Equivalence of actions in the confined phase}
The result in 4 dimensions is even more profound and interesting in the following sense. Not just that the infrared behaviour of the  theory with the quadratic gauge on the $\mathbb{S}^4$  and in the $\mathbb{R}^4$ is similar (Abelian dominance is observed in both spaces), the effective actions on the $\mathbb{S}^4$  and in the $\mathbb{R}^4$  in the corresponding ghost condensed phases also turn out to be  identical  as we see
now. 
To prove it, we need to know some relations of quantities on sphere with corresponding quantities in the Euclidean space  which are as follows \cite{5a'}
\be
 d \Omega& =& (\frac{1 + x^2}{2})^{-4} \ d^4 x, \nn\\
  \bar{F}_{\alpha \beta \gamma }^a \bar{F}_{\alpha \beta \gamma }^a &=& 3  (\frac{1 + x^2}{2})^{4} F^{\mu \nu a} F_{\mu \nu}^a,\\
  \bar{A}^a_\beta  \bar{A}_\beta^b (r)&=& (\frac{1 + x^2}{2})^{2}  {A}^a_\mu  {A}_\mu^b(x),\nn\\
  \bar{B}^a(r) &=&   (\frac{1 + x^2}{2})^{2} B^a(x).\nn
 \ee
 The first three of above relations can all be derived from the mappings given earlier whereas last one is intuitive from the dimensionality of fields.  We keep in mind the relation between masses on sphere and Euclidean space in Eq.\eqref{mr}.   Putting these relations along with  Eq.\eqref{mr}   in  the  effective action in the confined phase on $\mathbb{S}^4$ in Eq.\eqref{''}, we see that conformal factor, $\frac{1 + x^2}{2}$ exactly cancels and we get the  following term
 \be
 S_{eff} =   \int d^4 x [-\frac{1}{4}F^{\mu \nu a} F_{\mu \nu}^a + \frac{\xi}{2} (B^a)^2+ B^a {A}^a_\mu  {A}_\mu^a +  m_{a}^2 {A}^a_\mu {A}_\mu^a].
 \ee   The action on the right side is 4-dim Euclidean action in the ghost condensed phase~\cite{6}. The last term of this action  reaffirms that $m_a^2$ is square of mass of a gluon in the flat space.   Thus, it is proved that the  effective action in the confined phase on $\mathbb{S}^4$  and that in the 4-dim Euclidean space are identical.  The equality above is particular to 4 dimensions only.
 
\section{conclusion}
Infrared behavior of QCD on 4-sphere  and on a hypersphere in general has never been examined.   We investigated the infrared regime in the  quadratic gauge fixed theory  in the physical compact space   by defining it on 4-sphere. We observed that the inherent transversality condition plays a crucial role which gets rid of  the undesired  ghost term. One of the remaining ghost terms leads to Abelian dominance  on   $\mathbb{S}^{4}$ in the ghost condensed phase which signals the  presence of confinement on  $\mathbb{S}^{4}$ and in the general compact space.  
  The compactness of $\mathbb{S}^4$  does not affect  the infrared regime in  the present theory on  $\mathbb{S}^4$ in Eq.\eqref{28} from that in the quadratic gauge fixed theory in $\mathbb{R}^4$ but it  influences mass of an off-diagonal gluon on $\mathbb{S}^4$  to become position dependent in a way that  on the particular horizontal cross section of $\mathbb{S}^4$, the $\mathbb{S}^3$     mass remains same and it varies on parallel horizontal cross sections.    
  The equivalence between  the  effective actions  in the confined phases on $\mathbb{S}^4$  and  in the 4-dim Euclidean space was proved at the end.    The theory is also FFBRST compatible as it is BRST invariant. This will be particularly interesting area of further study as FFBRST has not been explored in the curved geometry.

\end{document}